\documentclass[aps,prl,superscriptaddress,twocolumn,tightenlines]{revtex4}

\usepackage {graphicx}
\usepackage {amsmath}
\usepackage {amsfonts}
\usepackage {amssymb}
\usepackage {mathrsfs}

\usepackage{epstopdf}

\newcommand {\be}{\begin{eqnarray}}
\newcommand {\ee}{\end{eqnarray}}

\begin{document}

\title {Observing the fluctuating stripes in high $T_c$
superconductors}

\author {V.\ Cvetkovic}
\email {vladimir@lorentz.leidenuniv.nl}
\affiliation {Instituut Lorentz voor de theoretische natuurkunde, 
Universiteit Leiden, P.O.\ Box 9506, 2300 RA Leiden, 
The Netherlands}
\author {Z.\ Nussinov}
\affiliation {Dept.\ of Physics, Washington University, St.\ Louis, MO 63160}
\author {S.\ Mukhin}
\affiliation {Theoretical Physics Department, Moscow Instituut for Steel \& alloys,
119991 Moscow, Russia}
\author {J.\ Zaanen}
\affiliation {Instituut Lorentz voor de theoretische natuurkunde, 
Universiteit Leiden, P.O.\ Box 9506, 2300 RA Leiden, 
The Netherlands}

\date {\today}

\begin{abstract}
Resting on the gauge theory of topological quantum 
melting in 2+1 dimensions, we predict that a superconductor
characterized by crystalline correlations on a length
scale large compared to the lattice carries a new collective mode:
the massive shear photon. This mode is visible in the
electrodynamic response and the ramification of our
theory is that electron energy loss spectroscopy
can be employed to prove or disprove the existence
of dynamical stripes in cuprate superconductors. 
\end{abstract}

\maketitle

The explanation of conventional superconductivity by the BCS
theory  belongs to the highlights of twentieth century physics.
However, it appears that BCS is too primitive to address 
the high $T_c$ superconductivity found in copper oxides. 
It is well established \cite{tranquada1,davis,abbamonte} 
that the superconducting liquid found in the cuprates is
in competition with the so-called stripe phase \cite{zagu}, 
which can be viewed as a 
special kind of electron crystal. By studying the behavior of the
spin \cite{tranquada1,kivelson,tranquada2} and phonon \cite{reznik} 
system with inelastic neutron scattering an intriguing case emerged that these stripes survive
in the form of transient correlations in the liquid  up to rather large length ($\sim 10$ nm)
and time ($\sim 10$ meV) scales. This affair is still quite 
controversial because the data can also be explained in
other ways \cite{hinkov,rpa1,rpa2} and there is a need for a unique, experimentally
accessible signature for the `quantum stripes' \cite{ZaanenNV}.
In this Letter we present such a feature which we
identified by studying the quantum field theory describing the 
topological quantum melting  of
crystals \cite{ZMN,kleinza,thesis,quantumsmectic}.
We claim that there should be
an extra propagating mode in the `nearly ordered'
superconductor, reflecting 
directly the presence of the transient stripe correlations. 

One can already anticipate the presence of the extra mode by
a simple quantum hydrodynamical consideration. 
Shear rigidity, the ability of the medium to 
respond reactively on a force acting in opposite directions
on opposite sides of the medium, 
is uniquely associated with {\em translational} symmetry breaking,
i.e., crystalline order. In a classical fluid one finds a viscous
response on shear, but in the dissipationless 
superfluids there is no  response to shear at all: these carry
only compressional rigidity, turning into magnetic (Meissner)
and electrical screening in the superconductor \cite {WenZee}. 
However, the `nearly ordered' superconductor should still have
the capacity to mediate shear forces at short distances and
in analogy with the London penetration
depth associated with magnetic forces, we call the 
characteristic length over which shear sources can make their
presence felt in the medium the `shear 
penetration depth' $\lambda_S$. The natural
velocity scale of reactive shear corresponds
to the transversal phonon velocity of the crystal $c_T$, and 
assuming a coherent quantum dynamics, one anticipates the
presence of  `photon-like' excitation in the system carrying
the shear stress, characterized by an energy
gap at zero momentum, $\Omega = c_T / \lambda_S$: the `massive shear
photon'. We claim that such a mode has to exist in high
Tc superconductors when the dynamical stripe interpretation is
correct, which implies that $\lambda_S \gg a$, the lattice constant.
Moreover,  a controlled mathematical description is available in this
limit \cite{ZMN,kleinza,thesis,quantumsmectic}, based
on modern developments in quantum field theory \cite{KleinertBookII,bais}, making it possible
to arrive at a detailed, quantitative prediction for an observable
quantity: the massive shear photon should give rise to a pole in
the electron-energy loss function at small momenta and energies (Fig.\ 1),
revealing its identity through the characteristic dependence of its 
pole strength on momentum (Fig.\ 1, inset).
 
The applicability of the field theory
requires some other conditions to be satisfied which are
in fact implied by the assumption of a large crystalline correlation
length.  This means that the system has to be
close to a continuous quantum phase transition
between the crystal and the superconductor, and this 
implies in turn: (a) The liquid and the solid have to be
microscopically similar and since the superconductor is a 
bosonic entity the crystal is also formed from bosons. It is
about preformed pairs which can either form stripes or the
superconductor \cite{scalapino}. (b) 
The transition from the crystal to the superconductor
is first order, and to make possible a large shear length  
it appears necessary to assume that
the superconductor is at the same time a quantum liquid 
crystal of either the smectic or nematic kind \cite{quliqcrys}. 
It turns out that with regard to the shear stress photon it does
not make much of a difference \cite{thesis} and we will present 
here results for the simpler nematic superconductor. 

The theory can be viewed as the quantum generalization of
the famous Nelson-Halperin-Young theory \cite{NHY} 
of classical melting in two dimensions, based on the notion 
that the liquid crystal can be viewed as a crystal where
topological defects (dislocations) have proliferated. 
The quantum version rests on a Kramers-Wannier duality arising
in elasticity theory, first explored by Kleinert
in the 1980's in the context of the classical problem in
three dimensions \cite{KleinertBookII}, 
and only recently implemented on the quantum
level \cite{ZMN}. This earlier work contained some technical
flaws (ignoring the relativistic character of the dislocation
condensate, a faulty gauge fix) obscuring the view on the
electrodynamic response. We present here
a summary of the main steps of the correct derivation, and we
refer for further details to Ref.'s\ \cite{quantumsmectic,thesis,hexaticlong}.

In the `orderly' limit $\lambda_S / a \rightarrow \infty$, the
constituent bosons of the gaseous limit are no longer relevant
and instead the work is done by the collective degrees of freedom:
the phonons and the topological defects. The latter correspond 
to dislocations and disclinations, restoring the translational and
rotational symmetry, respectively. In
this language, liquid crystals  are states 
where dislocations  have proliferated, restoring 
translations, while disclinations  are still massive
excitations such that rotational symmetry remains broken
\cite{quantumsmectic,bais}. The phonons of a (`Wigner') crystal
of charged bosons in 2+1 Euclidean space-time dimensions
are described by the following Lagrangian density,
following \cite{ZMN} from the gradient expansion in terms of
the displacement fields $u^a$,
\be
  {\mathscr L}_0 = \tfrac 1 2 \partial_\mu u^a C_{\mu \nu a b} \partial_\nu u^b
  + i {\cal A}_\mu^a \partial_\mu u^a + \tfrac 1 4 F_{\mu \nu} F^{\mu \nu},
  \label {L0}
\ee
where $C_{\mu \nu a b}$ is a short hand for the tensor of
elastic moduli, now including the kinetic energy
$(\rho / 2) (\partial_{\tau} u^a)^2$ (Greek indices
label space-imaginary time directions $\sim x,y,\tau$, 
Latin indices space directions $\sim x, y$; notice that 
$C_{\tau \tau a b} = \rho \delta^{ab}$ where $\rho$ is
the mass density). For simplicity
we limit ourselves to isotropic elasticity characterized
by just a shear and compression modulus, $\mu$ and $\kappa$
respectively, which are related by the Poisson ratio
$ \kappa = \mu ( 1 + \nu)/( 1- \nu)$. This defines the velocities
$c_T = \sqrt{ \mu / \rho}$, $c_L = c_T \sqrt{2 / (1-\nu)}$,
$c_{\kappa} = c_T \sqrt{ ( 1+ \nu) / ( 1- \nu)}$, corresponding to
the transversal and longitudinal phonon velocities of the crystal, while
$c_{\kappa}$ is the compressional (real sound) velocity.
The last term is the usual Maxwell term
describing the electromagnetic fields while the electromagnetic
vector potentials ($A_{\mu}$)  couple to the elastic strains 
via the effective combination
${\mathcal A}_\mu^a = (n_e e^*) \left \lbrack A_\tau \delta_{\mu a} -
A_a \delta_{\tau \mu} \right \rbrack$ ($e^*$ is the microscopic
electrical charge, $n_e$ is the density of charged particles).
The scales associated with the electromagnetism are the plasma 
frequency $\omega_p = \sqrt{4 \pi n_e e^2 / \rho}$  and the Debye
electrical screening length $\lambda_e = c_T / \omega_p$ and
Debye momentum $q_e = 1/ \lambda_e$.

The crucial insight, due to Kleinert, is to turn this
into a `stress gauge theory' \cite {KleinertBookII}. The first step is
to use
strain-stress duality, 
such that Eq.\ (\ref{L0}) becomes,
\begin{eqnarray}
  {\mathscr L}_{dual}  =  \tfrac{1}{2} \left \lbrack
  \tfrac{ (\sigma^a_a )^2 }{\kappa} + 
   \tfrac{ (\sigma^x_x - \sigma^y_y )^2 + (\sigma^x_y + \sigma^y_x )^2}{4\mu}
   + \tfrac{ (\sigma^a_{\tau} )^2 }{\rho} \right \rbrack + \nonumber \\
     i {\mathcal A}_\mu^a C_{\mu \nu a b}^{-1}
  \sigma_\mu^a + \tfrac 1 2 {\mathcal A}_\mu^a C_{\mu \nu a b}^{-1}
  {\mathcal A}_\nu^b + \tfrac 1 4 F_{\mu \nu} F^{\mu \nu}, 
  \label {Ldual}
\end{eqnarray}
in terms of the stress fields 
$\sigma^a_{\mu}$, with the $\sigma^{a}_{\tau}$'s
having the status of canonical momenta. Notice that
the antisymmetric components of the spatial stress vanish,
$\sigma^x_y - \sigma^y_x = 0$ (Ehrenfest constraint).
The key is that this is augmented by the law expressing
the conservation of stress $\partial_{\mu} \sigma^a_{\mu} =0$
and this can be imposed in 2+1D by expressing the stresses
in terms of the stress gauge fields,
\be
  \sigma_\mu^a = \epsilon_{\mu \nu \rho} \partial_\nu B_\rho^a.
 \label {sigma}
\ee
The magic of these stress-gauge fields is, that like in
vortex duality \cite {XYCvetkovic}, 
the sources of these stress gauge fields
correspond to the non-integrable displacement field
configurations \cite {KleinertBookII},
\begin{equation}
   {\mathscr L}_{disl}  =  i B^a_{\mu} J^a_{\mu}, \; \; \;
  J^a_{\mu}  =  \epsilon_{\mu \nu \lambda} 
  \partial_{\nu} \partial_{\lambda} u^a,
  \label{dislcurrents}
\end{equation}
corresponding to the  dislocation currents which can be
factorized as $J_\mu^a = b^a {\mathscr J}_\mu$:
world-lines of dislocations  with
Burgers vector ${\bf b} = (b_x, b_y)$. This implies in turn that
$B_\mu^a J_\mu^a \to (b^a B_\mu^a) {\mathscr J}_\mu$. 
Hence, we arrive at a description of elasticity
which is a close sibling of electromagnetism: the stress
gauge fields (`stress photons') express the capacity of the 
elastic medium to propagate elastic forces, and the dislocations
take the role of charged sources.

In the 2+1 space-time dimensions of relevance to cuprates, 
the miracle is that the quantum-nematic crystal can be 
equally well viewed as a `dual superconductor'  \cite{ZMN,kleinza}, 
in close analogy with the 
superfluid--superconductor duality in 2+1 dimensional phase
dynamics \cite{Fisher,Hove,XYCvetkovic}. In the path-integral
language, the `dual nematic shear superconductor' corresponds
to a tangle of dislocation worldlines in 
space-time. As we explained elsewhere \cite {quantumsmectic},
these describe either smectic or nematic quantum liquid crystals
depending on the orientational ordering of the Burgers vectors:
in the smectic, dislocations condense
with their Burgers vectors oriented in one particular direction
while in the nematic they condense with equal
probability for their Burgers vectors to be oriented along
all allowed directions \cite {bais}. As we are
dealing with isotropic elasticity, we focus on
the quantum hexatic, derived from the hexagonal crystal
where the Burgers vectors are equally distributed along the
six directions associated with the rotational $D_6$ symmetry.

The dislocation condensate is gauged by the stress gauge fields, and
Eq.\ (\ref{dislcurrents}) implies a
covariant derivative structure of the form
$| (\partial_{\mu} - i b^a B^a_{\mu} ) \Psi |^2$ where $\Psi$
is the order parameter of the dislocation condensate; as we
overlooked in the earlier work \cite{ZMN}, it is vital to keep here 
the time derivatives given the relativistic nature of the dislocation
condensate \cite{XYCvetkovic}. Elsewhere, we will present in detail 
the averaging procedure leading to the correct Higgs term in 
the stress-sector \cite {thesis,hexaticlong}. We find a `bare' 
Higgs term
for the  quantum hexatic ${\mathscr L}_{H, bare} = \tfrac 1 4
|\Psi_0|^2 B_\mu^a B_\mu^a$, where $\Psi_0$ is the expectation value
of the disorder field. Both the
glide \cite {glide} and Ehrenfest constraints have to be
imposed on the bare Higgs term. The glide constraint is implemented 
via a Lagrange multiplier term ${\mathscr L}_{glide} = i \lambda (J_x^y - J_y^x)$.
This operation
removes the compressional stress $\sigma_x^x + \sigma_y^y$
from the Higgs term, having also the effect
of turning \cite{ZMN, hexaticlong, thesis} the
second sound velocity of the condensate into the
`glide velocity' $c_g = c_T / \sqrt 2$. Subsequently, the
Ehrenfest constrained is implemented by eliminating the
antisymmetric stresses $\sigma^y_x - \sigma^x_y$ from the
Higgs term. It is particularly convenient to represent
the resulting Higgs term in a gauge invariant way \cite {kleinza},
\be
  {\mathscr L}_{H} = \tfrac 1 2 \tfrac {\Omega^2}{4 \mu} \left \lbrack \tfrac
  {2 (\sigma_x^x - \sigma_y^y)^2}{(\partial_\tau)^2 +
  {c_g^2} (\partial_x)^2} + \tfrac {(\sigma^x_y + \sigma^y_x)^2
  (1 + {c_g^2 (\partial_x)^2 } / {(\partial_\tau)^2})}
  {(\partial_\tau)^2 + {c_T^2} (\partial_x)^2} \right \rbrack ,
  \label {LH}
\ee
where we defined the `shear Higgs mass' as $\Omega = | \Psi_0 | \sqrt \mu$.

This result is revealing: by counting derivatives it is
immediately obvious that both the longitudinal
($\sigma^x_x - \sigma^y_y$) and transversal
($\sigma^y_x + \sigma^x_y$) shear stresses acquire a Higgs mass $\Omega$,
while compressional stress is left unaffected. 

\begin{figure} 
\begin {center}
\includegraphics[width=0.48\textwidth]{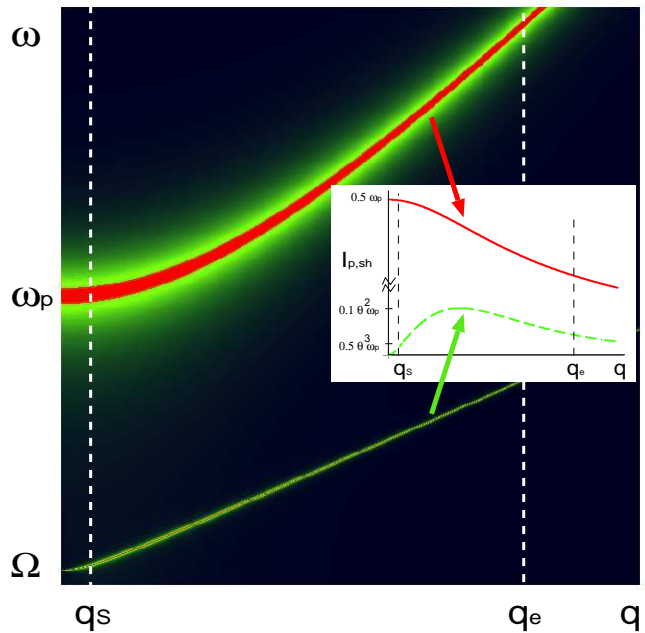}
\end {center}

\caption{The electromagnetic absorptions in the `nearly ordered'
superconductor as seen by electron-energy loss spectroscopy
(${\rm Im} \lbrack 1 / \varepsilon_L ( q, \omega) \rbrack$) as function
of frequency $\omega$ and momentum $q$. In this example, we take a typical 
Poisson ration $\nu = 0.28$, characterizing the `background' Wigner 
crystal, while the `shear Higgs mass' $\Omega = 0.05 \omega_p$ and 
shear penetration depth $\lambda_S = 1 / q_S = c_T / \Omega$ are
taken to be representative for the situation cuprate superconductors
($\omega_p \simeq 1$ eV, $\Omega \simeq 50$ meV, $\lambda_S \simeq 10$
nm). Besides the strong plasmon pole dominating the long wavelength
dielectric response, we also find a weak absorption which corresponds
to the massive shear photon giving away the presence of the `dual
shear superconductor', to be regarded as the unique fingerprint
of a superconductor characterized by transient translational order
extending over distances large compared to the lattice constant.
In the inset, the weights of the plasmon (solid red, $I_p$) and the  
shear photon (dashed green, $I_{sh}$) in the electron energy 
loss spectrum as function of momentum are given. The latter can be 
used
to isolate the shear photon in 
the experimental data. At long wavelengths ($q < q_S$) 
$I_{sh}$ grows quadratically with momentum, with a prefactor
set by the ratio of the shear mass and the plasmon
frequency $\theta = \Omega/ \omega_p$, while it reaches a
maximum at at intermediate wavelengths ($q_S < q < q_e$).}    
 \label {FigEELSPlots}
\end{figure}

Ignoring the EM fields, the elastic response of the neutral
quantum hexatic can be straightforwardly calculated from
Eqs.\ (\ref{Ldual}, \ref {LH}) yielding similar results for
the mode spectrum as we discussed for the smectic \cite {quantumsmectic}.
Let us now focus on the electromagnetic response, the only way
to probe the system from the outside given that we deal with 
electrons. According to standard linear response theory, the 
electrodynamic response is governed by the dielectric tensor, 
which is completely parametrized in terms of transversal and 
longitudinal dielectric functions. For isotropic
media, these can be expressed
in terms of diagonal elements of the photon self-energy
$\Pi$ as \cite {Mahan} (with $q$ the momentum and
$\omega_n$ the Matsubara frequencies),
\begin{eqnarray}
  \hat \varepsilon_L (q, \omega_n) = 1 - 
  \tfrac {\Pi_\tau (q, \omega_n)} {q^2}, \;
  \hat \varepsilon_T (q, \omega_n)  = 1 - 
  \tfrac {\Pi_T (q, \omega_n)} {\omega^2_n},
\label{dielectric}
\end{eqnarray}
in the Coulomb gauge for the EM fields, $\partial_a A_a = 0$.
We observe that the total dual action Eq.\ (\ref {Ldual}, \ref {LH})
can schematically be rewritten as,
\be
  {\mathscr L}_{dual} = \tfrac 1 2 B^a_{\mu} 
({\mathcal G})_{\mu \nu a b}^{-1}
  B^b_{\nu} + i B^a_{\mu} g_{a, \mu \nu} A_\nu  + \nonumber \\
  \tfrac 1 4 F_{\mu \nu} F^{\mu \nu} -
  \tfrac 1 2 A_\mu (\Pi^{bare})_{\mu \nu} A_\nu. \label {Ldual2}
\ee
where all the physics of the stress sector is lumped together
in the fully dressed stress gauge field propagator 
$({\mathcal G})_{\mu \nu a b}$. 
Given that the coupling between stress and EM gauge fields
is linear, the photon self-energies are easily derived 
by straightforwardly integrating
out the stress photons from Eq.\ (\ref{Ldual2}),   
\be
  - \Pi_{\mu \nu} (q, \omega_n) = - \Pi_{\mu \nu}^{bare} - g^*_{a,\kappa,\mu}
 ({\mathcal G} (q, \omega_n))_{\kappa \kappa' a b}   g_{b,\kappa',\nu}. 
\label {Pi}
\ee
The coupling constants $g_{a, \mu \nu}$ and the $\Pi^{bare}$ are
tabulated in Ref.\ \cite {thesis}. 
It is a straightforward exercise
to obtain explicit expressions for the dielectric functions
Eqs.\ (\ref{dielectric}). 

As we already noticed in the earlier work \cite{ZMN}, from the transverse
electromagnetic response it follows that the quantum hexatic is also
an electromagnetic Meissner state: this should be surprising because
this superconductivity does not involve off-diagonal long range order
of the constituent bosons and the presence of the dual dislocation state
turns out to be a sufficient condition. In the remainder of this paper, 
we will focus on the longitudinal electromagnetic response for which
\begin{equation}
  \varepsilon_L 
  = 1 +  \frac { \omega^2_p (c_g^2 q^2 + \Omega^2 - \omega^2)}
{(c_g^2 q^2- \omega^2) ( c_L^2 q^2 - \omega^2) + 
\Omega^2 (c_\kappa^2 q^2 - \omega^2)}. \label {epsL}
\end{equation}
This is our main result: the longitudinal response is at least in 
principle accessible by experiment at the finite momenta which are 
required to see the influences of the transient order: all that is needed 
is a measurement of the electron energy loss spectrum corresponding 
to ${\rm Im} \lbrack 1 /  \varepsilon_L (q, \omega) \rbrack$ in the relevant 
kinematic regime. 
The existence of two modes is evident from the poles. 
In the limit $\lambda_S/a \rightarrow \infty$,
this expression has a universal status for small momenta; at larger
momenta one has to account also for damping by perturbative mode couplings, single particle excitations, etcetera. 
In Fig.\ 1, we
show an example of the loss-spectrum which could be representative for
cuprates assuming that the characteristic disorder scales revealed by
the measurements of the spin fluctuations are rooted in the charge
fluctuations discussed here: as compared to the plasma frequency
$\sim 1$ eV, the shear Higgs mass $\Omega$ should be small, 
($\sim 10^{-2}$ eV), while the characteristic length scale parametrized
by $\lambda_S$ should be $\sim 10$ nm), consistent with a typical
electronic velocity $c_T \simeq 1$ eV{\AA}. The bottom line is that
besides the intense plasmon one finds a weak feature at low energy 
in the loss spectrum, corresponding to the massive shear photon.

The weight of the shear photon has a non-trivial dependence on both
the ratio $\theta = \Omega / \omega_p$ as well as on the momentum $q$.
This is easily computed from Eq.\ (\ref {epsL}) and the results 
are summarized in inset of Fig.\ 1. In the
`liquid' long wavelength regime $q < q_S$, the weight of
the shear photon $I_{sh} (q) = \frac{\theta (1 -\nu)}
{4 (1 - \theta^2)} \; q^2 + O(q^4)$.\ $I_{sh}$ has
a maximum at intermediate momenta $q_{max} = 
\sqrt{2(1-\nu)/(3(3+\nu))} \sim 0.4 q_e$ 
where the shear photon acquires
a weight relative to the plasmon $(I_{sh} (q_{max}) / I_p)
\simeq (3\sqrt{3(1-\nu)})/(16\sqrt{3+\nu})\; \theta^2$ which
we expect to be of order $0.1\%$ in cuprates given that 
$\theta \simeq 0.05$: the electromagnetic weight of the shear 
photon is expected to be very small. This momentum and $\theta$ 
dependence of the electromagnetic strength of the shear photon can
however be taken as its fingerprint, since it is rooted in the
mechanism through which the shear photon becomes electromagnetically
active. Both shear stress and dislocations forming the building
blocks for the massive shear photon do not carry volume and are therefore
electrically neutral. However, in the Wigner crystal 
and in the liquid, at distances small compared to $\lambda_S$, 
the plasmon carries like longitudinal phonon shear components at finite momenta. 
These shear components have to be `removed' from the plasmon, turning it into the purely compressional mode of the liquid at distances large compared to
$\lambda_s$. This causes a linear mode coupling between the plasmon and
the shear photon, with the latter `stealing' some electromagnetic weight
from the former. This explains the strong dependence on $\theta$, the
characteristic maximum, and the vanishing of the
weight at $q \rightarrow  0$ as even in the Wigner crystal, the
plasmon turns into a purely compressional mode in the long wavelength
limit.

What has to be done to detect the shear photon? It is clear
that one needs a very high resolution measurement of the 
loss function, given its low energy ($\sim 10$ meV) and very
small spectral weight. Optical methods are not suited because
these measure at momenta which can be regarded as infinitesimal
even on the scale of $q_S$. Obviously, the easy way is to measure directly
the loss function in this kinematic regime. This requires
a millielectronvolt energy, and nanometer spatial resolution.
Although such instruments are not available right now, it just
seems to involve a relatively minor engineering effort to get
into this regime: reflection (low energy) EELS has the
resolution but for unknown reasons there are problems measuring
the electronic response \cite{Sawatzky}; 
transmission (high energy) EELS is
limited to a resolution of order of $\sim$ 100 meV but there 
seems much room for improvement using modern electron optics. 
The strongest contender appears to be resonant inelastic 
soft X-Ray scattering, where at present a major
instrumental development program is unfolding aimed at reaching meV 
resolution.

\section {Acknowledgments}

This work was supported by the Netherlands
foundation for fundamental research  of Matter (FOM).

\bibliographystyle{apsrev}

\end {document}